\documentclass[aps,prb,reprint,amsmath,superscriptaddress,amssymb,graphicx]{revtex4-1}
\usepackage{mathrsfs}
\usepackage{graphicx,amsmath}
\usepackage{color}
\usepackage{epstopdf}
\begin{document}

\title{Anomalous spin Nernst effect in Weyl semimetals}

\author{Ning-Xuan Yang}
\affiliation{International Center for Quantum Materials, School of Physics, Peking University, Beijing 100871, China}

\author{Yan-Feng Zhou}
\affiliation{International Center for Quantum Materials, School of Physics, Peking University, Beijing 100871, China}

\author{Zhe Hou}
\affiliation{International Center for Quantum Materials, School of Physics, Peking University, Beijing 100871, China}

\author{Qing-Feng Sun}
\email[]{sunqf@pku.edu.cn}
\affiliation{International Center for Quantum Materials, School of Physics, Peking University, Beijing 100871, China}
\affiliation{Beijing Academy of Quantum Information Sciences,
West Bld.\#3,No.10 Xibeiwang East Rd., Haidian District, Beijing 100193,China}
\affiliation{Collaborative Innovation Center of Quantum Matter, Beijing 100871, China}

\date{\today}

\begin{abstract}
The spin Nernst effect describes a transverse spin current induced by the
longitudinal thermal gradient in a system with the spin-orbit coupling.
Here we study the spin Nernst effect in a mesoscopic four-terminal cross-bar Weyl semimetal
device under a perpendicular magnetic field.
By using the tight-binding Hamiltonian combining with the nonequilibrium Green's function method,
the three elements of the spin current in the transverse leads
and then spin Nernst coefficients are obtained.
The results show that the spin Nernst effect in the Weyl semimetal
has the essential difference with the traditional one:
The $z$ direction spin currents
is zero without the magnetic field while it appears under the magnetic field,
and the $x$ and $y$ direction spin currents in the two transverse leads
flows out or flows in together,
in contrary to the traditional spin Nernst effect, in which the spin current is induced by the spin-orbit coupling
and flows out from one lead and flows in on the other.
So we call it the anomalous spin Nernst effect.
In addition, we show that the Weyl semimetals have the center-reversal-type symmetry, the
mirror-reversal-type symmetry and the electron-hole-type symmetry,
which lead to the spin Nernst coefficients being odd function or even function
of the Fermi energy, the magnetic field and the transverse terminals.
Moreover, the spin Nernst effect in the Weyl semimetals are
strongly anisotropic and its coefficients are strongly dependent
on both the direction of thermal gradient and
the direction of the transverse lead connection.
Three non-equivalent connection modes ($x$-$z$, $z$-$x$ and $x$-$y$ modes) are studied in detail,
and the spin Nernst coefficients for three different modes exhibit very different behaviors.
These strongly anisotropic behaviors of the spin Nernst effect
can be used as the characterization of magnetic Weyl semimetals.
\end{abstract}


\maketitle

\section{\label{sec1} Introduction}

Weyl semimetals (WSMs) are a novel topological quantum state in condensed matter\cite{LvBQ,XuSY,WanX,WengH,Jiang,Jiangqd,ChenCZ1},
which are characterized by the existence of a set of linear-dispersive
band-touching points, known as the Weyl nodes.
The Weyl nodes always appear in pairs, of which the quasiparticles carry opposite chirality.
In the momentum space the Weyl node acts like a source or drain of the Berry curvature,
resulting in the Fermi arc surface states which take a form of
a finite segment terminated at the Weyl nodes.\cite{XuSY1,Huang,XuSY2,Zhang}
Besides, WSMs also manifest lots of exotic properties in quantum transport,
such as chiral anomaly induced negative magnetoresistance,\cite{Hosur,Burkov,Xiong,LuHZ1,LiY,Jia,ChenCZ2}
chiral magnetic effect,\cite{SonDT,Zyuzin1} weak anti-localization,\cite{LuHZ2}
double Andreev reflections,\cite{HouZ} etc.
Due to their unique gapless bulk states, the Fermi arc surface states
and special transport properties, WSMs have attracted significant
attention.\cite{Lundgren,ChenQ,Igarashi,Ramak,Ominato,Zyuzin2,McCormick,Gorbar}
For example, Lundgren and Chen \textit{et al.} investigate
the electronic contribution to the thermal conductivity
and the thermopower of Weyl and Dirac semimetals using the Boltzmann equation.\cite{Lundgren,ChenQ}
Igarashi \textit{et al.} theoretically study electronic transport
in the WSM nanowires under magnetic fields,\cite{Igarashi}
and demonstrate that the interplay between the Fermi-arc surface states
and the bulk Landau levels plays a crucial role in the magnetotransport.

The spin Nernst effect refers to a transverse spin current caused
by the longitudinal temperature gradient in a system with spin-orbit coupling.\cite{ChengS}
Recently, this effect has been observed for the first time
in a six-terminal Hall-bar Platinum thin film system\cite{Meyer},
which makes it one of the most exciting subjects in spintronics.
To date, more and more research groups are studying the spin Nernst effect
in various systems of different materials.
For example, Sheng \textit{et al.}\cite{ShengP} observed the spin Nernst effect
in W/CoFeB/MgO heterostructures, and Bose \textit{et al.}\cite{Bose} observed
the heat current to spin current conversion in non-magnetic Platinum
by the spin Nernst effect at room temperature.
Similar to the Nernst coefficient being more sensitive to the details of
the density of states than the Hall conductance,\cite{Xingy,YangNX}
the spin Nernst coefficient is more sensitive to the details of
the spin density of states of the system than the spin Hall conductance.
The spin Nernst effect also offers a possibility of controlling
the electron spin current in spintronics applications.\cite{Meyer,ShengP,Bose}
Furthermore, the non-dissipative pure spin current not only deepens
the understanding of the phenomenon of non-dissipative quantum transport,
but also contributes to the development of novel low power-consumption
nanoscale spintronic devices.\cite{Murakami,LiuL}

Due to its potential application in spintronics,
the spin Nernst phenomenon has long been the focus of theoretical research
and has attracted wide attention of researchers.\cite{ChengS,TauberK,Rothe,Wimmer,Dyrda,LiuX}
Early in 2008, Cheng and Sun\cite{ChengS} first theoretically studied
the spin Nernst effect in a two-dimensional electron gas system
with spin-orbit coupling under a perpendicular magnetic field.
It was found that the spin-orbit coupling can lead to splitting of the Nernst peak,
and the spin Nernst coefficient increases with the spin-orbit coupling strength,
but weakens with the increase of the magnetic field.
Thereafter, a lot of theoretical works have studied the spin Nernst effect
in various systems in depth.\cite{TauberK,Rothe,Wimmer}
Tauber \textit{et al.} calculated the influence of impurities
on the Nernst effect,\cite{TauberK} and found that the direction and magnitude of
spin current can be modulated by changing the type of impurity.
Rothe \textit{et al.} investigated the spin-dependent thermoelectric transport
in quantum spin Hall insulators based on HgTe/CdTe quantum wells
in the absence of magnetic fields.\cite{Rothe}
It was found that the oscillatory character of the spin Nernst coefficients
in the bulk gap were caused by the finite overlap of
the edge states from opposite sample boundaries.
Wimmer \textit{et al.} presented the first-principles description of
the spin Nernst effect based on the Kubo-St$\check{r}$eda formalism,\cite{Wimmer}
and used this method to study the spin Nernst effect of diluted and concentrated alloys.

In WSMs, spin-momentum locking correlates the spin direction with the orbital motion of electrons.
The spin direction is parallel or anti-parallel with the momentum direction
due to the well conserved chirality of each Weyl node,
making WSMs a good platform for investigating the spin transport.
However, up to now, no investigations of the spin Nernst effect in WSMs have been reported,
although the Nernst effect in WSMs were investigated by some recent works.\cite{Watzman,Sharma,Ferreiros,Caglieris,Noky,Chernodub,SahaS}

In addition, we also note that the spin current is a tensor,\cite{spinc1,spinc2,spinc3}
which has $3 \times 3=9$ elements, describing the direction of the electron motion
and the direction of the spin, respectively.
While in a lead, electrons have to move along the lead,
but its spin direction may still be in the $x$, $y$ and $z$ directions.
In this case, the spin current has three non-zero elements, $I_{sx}$, $I_{sy}$ and $I_{sz}$.
Here $I_{s\mathrm{i}}$ ($\mathrm{i}=x$, $y$ and $z$) represents an electron
moving along the lead with its spin in the $\mathrm{i}$ direction.
However, all previous studies investigated only the element $I_{sz}$ in the spin Nernst effect.
The spin currents $I_{sy}$ and $I_{sz}$ have never been studied.

\begin{figure}
\includegraphics[scale=0.85]{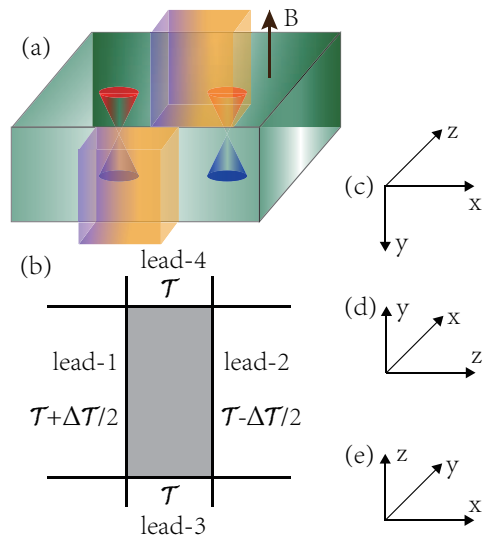}
\caption{
(a) and (b) are the schematic cubic diagram
and the top view of system, which consists of a rectangular center WSM region
connected to four ideal semi-infinite leads. The lead-1 and lead-2 are the semi-infinite WSM,
and the lead-3 and lead-4 are the normal conductor.
For the WSM, it has a pair of Weyl nodes at $\textbf{K}_{\pm}=(0,0,\pm\pi/(2a))$.
A thermal gradient $\Delta\mathcal{T}$ is applied between the longitudinal lead-1 and lead-2,
and it will induce the spin current in the transverse lead-3 and lead-4.
(c), (d) and (e) represent the coordinate systems of three non-equivalent $x$-$z$, $z$-$x$ and $x$-$y$
connection modes, respectively. }
\end{figure}

In this paper, we carry out a theoretical study of the spin Nernst effect of WSMs
under the perpendicular magnetic fields by using the Landauer-B\"{u}ttiker formula combining
with the nonequilibrium Green's function method.
We consider a time-reversal symmetry breaking WSMs in a mesoscopic four-terminal cross-bar device.
The three elements of the spin Nernst coefficients are calculated at
different temperatures in three connection modes ($x$-$z$, $z$-$x$ and $x$-$y$ modes).
We find that the spin Nernst effect in the WSMs has the essential difference with
the traditional spin Nernst effect. So we call it the anomalous spin Nernst effect.
The anomalous behavior is that
1) the $z$ direction element of the spin Nernst coefficients
is zero at the zero magnetic field, and it appears at the presence of the magnetic field,
in contrary to the traditional one induced by the spin-orbit coupling,
and 2) the $x$ and $y$ direction elements of the spin currents in the two transverse leads
flows out or flows in together, which is essentially different with the traditional one, where
the spin current flows out from one lead and flows in on the other.
In addition, the spin Nernst coefficients show the strongly anisotropic characteristics in space.
For the $x$-$z$ and $z$-$x$ connection modes the spin Nernst coefficients show a series of peaks,
and the peak positions are independent the magnetic field,
but they strongly oscillate and are very sensitive to the magnetic field for the $x$-$y$ mode.
Moreover, through the analysis with both the continuous Hamiltonian and discrete Hamiltonian,
the center-reversal-type symmetry, the mirror-reversal-type symmetry and the electron-hole-type symmetry
are found, which lead to the spin Nernst coefficients being odd function or even function
of the Fermi energy, the magnetic field and the transverse terminals.

The rest of the paper is organized as follows.
In Sec.~\ref{sec2}, the effective tight-binding Hamiltonian is introduced,
and the formalisms for calculating the spin Nernst coefficients
$N_{sx}$, $N_{sy}$ and $N_{sz}$ are derived.
In Sec.~\ref{sec3}, Sec.~\ref{sec4} and Sec.~\ref{sec5},
we study the spin Nernst effect under the
$x$-$z$, $z$-$x$ and $x$-$y$ connection modes, respectively.
Finally, a brief summary is drawn in Sec.~\ref{sec6}.

\section{\label{sec2} Model and Methods}

Here we consider the time-reversal symmetry breaking WSMs under
a perpendicular magnetic field as shown in Fig.1(a).
In the momentum space with the zero magnetic field,
the Hamiltonian of the WSMs can be described as~\cite{Yang,Igarashi}:
\begin{eqnarray}\label{eq:1}
&H=2t \sigma_{z}(2+\cos{k_{z}a}-\cos{k_{x}a}-\cos{k_{y}a})&      \nonumber \\
   &+2t \sigma_{x}\sin{k_{x}a}+ 2t \sigma_{y}\sin{k_{y}a}.&
\end{eqnarray}
Here $\sigma_x$, $\sigma_y$ and $\sigma_z$ are the Pauli matrices,
and $t =\hbar v/2a$ with the Fermi velocity $v$ and the lattice constant $a$.

From Eq.(\ref{eq:1}), one can easily verify that there exists one pair of Weyl nodes at
$\textbf{K}_{\pm}=(0,0,\pm\pi/(2a))$ in the bulk Brillouin zone.
Due to the Weyl nodes being at the $k_z$ axis,
it leads to different properties between $z$ direction and $x$ ($y$) direction,
and the Hamiltonian of WSMs in Eq.(\ref{eq:1}) shows anisotropic.
Such anisotropy will result in the spin Nernst coefficient being related
to the direction of thermal gradient and direction of transverse lead connection.
We consider a WSM system consisting of a rectangular center WSM region
connected to four ideal semi-infinite leads, as shown at the schematic cubic diagram
in Fig.1(a) and the top view of system in Fig.1(b).
A longitudinal thermal gradient $\Delta\mathcal{T}$ is added between lead-1 and lead-2.
This thermal gradient induces a transverse spin current $I_{\mathrm{q}s}$
at the lead-3 and lead-4.

As mentioned above, WSMs are anisotropic and the spin Nernst effect
strongly depends on both the direction of thermal gradient
and direction of transverse lead connection.
There totally are six different connection modes, $x$-$z$, $x$-$y$,
$y$-$x$, $y$-$z$, $z$-$x$ and $z$-$y$, for the connection of the four leads.
Here $\mathrm{i}$-$\mathrm{j}$ mode ($\mathrm{i},\mathrm{j}=x, y, z$) represents
that thermal gradient is applied in the $\mathrm{i}$ direction
and the spin current is measured in the $\mathrm{j}$ direction,
i.e. the lead-1 and lead-2 connect the rectangular center WSM region at the $\mathrm{i}$ direction
and lead-3 and lead-4 are at the $\mathrm{j}$ direction.
However, $k_{x}$ and $k_{y}$ in Eq.(\ref{eq:1}) are equivalent
and only $k_{z}$ is special, this means that the $x$ and $y$ directions are equivalent,
which results in an equal spin Nernst coefficients in the two modes $x$-$z$ and $y$-$z$
($x$-$y$ and $y$-$x$, $z$-$x$ and $z$-$y$).
So there are only three non-equivalent cases of the six modes.
Thereafter, we consider three non-equivalent $x$-$z$, $z$-$x$ and $x$-$y$ modes
[see Fig.1(c,d,e)].
For example, for the $x$-$z$ mode, the longitudinal thermal gradient $\Delta\mathcal{T}$
is added in the $x$ direction, and the spin current is measured in the $z$ direction.
The width in the $y$ direction is assumed to be wide, the periodic boundary condition is used,
and the momentum $k_y$ is a good quantum number.
We also consider a magnetic field applied in the $y$ direction,
that is, perpendicular to the transport plane.
For the $z$-$x$ ($x$-$y$) mode, the width in the $y$ ($z$) direction is set to be wide
and the momentum $k_y$ ($k_z$) is a good quantum number.
The magnetic field applies in the $y$ ($z$) direction,
and it is perpendicular to the transport plane still.

Based on the Hamiltonian of Eq.(\ref{eq:1}), we here propose
the two-band tight-binding discrete model on a simple cubic lattice.
For $x$-$z$ and $z$-$x$ modes, the momentum $k_{y}$ is a good quantum number,
and the tight-binding lattice models can be written as
\begin{eqnarray}\label{eq:2}
&&H(k_{y})=\sum_{\mathbf{j}}[ c_{\mathbf{j} }^{\dag}T_{0}c_{\mathbf{j} }+ c_{\mathbf{j} }^{\dag}T_{x}c_{\mathbf{j}+\delta{x}}
      +c_{\mathbf{j}}^{\dag}T_{z}c_{\mathbf{j}+\delta{z}}+ \textrm{H.c.}], \nonumber   \\
&&T_0=4t\sigma_z+ 2t\sigma_y \sin{k_{y}a}- 2t\sigma_z \cos{k_{y}a}, \nonumber   \\
&&T_x=[-t\sigma_z- it\sigma_x]e^{-i(\phi^{\perp}_{\mathbf{j},\mathbf{j}+1})}, \nonumber   \\
&&T_z=t\sigma_z e^{-i(\phi^{\perp}_{\mathbf{j},\mathbf{j}+1})}.
\end{eqnarray}
For $x$-$y$ mode, the momentum $k_{z}$ is a good quantum number,
and the tight-binding lattice model is written as
\begin{eqnarray}\label{eq:3}
&&H(k_{z})=\sum_{\mathbf{j} }[ c_{\mathbf{j} }^{\dag}R_{0}c_{\mathbf{j} }+ c_{\mathbf{j} }^{\dag}R_{x}c_{\mathbf{j}+\delta{x}}
      +c_{\mathbf{j} }^{\dag}R_{y}c_{\mathbf{j}+\delta{y}}+ \textrm{H.c.}], \nonumber   \\
&&R_0=4t\sigma_z+ 2t\sigma_z \cos{k_{z}a},  \nonumber \\
&&R_x=[-t\sigma_z- it\sigma_x]e^{-i(\phi^{\perp}_{\mathbf{j},\mathbf{j}+1})},  \nonumber \\
&&R_y=-t\sigma_z- it\sigma_y,
\end{eqnarray}
where $a$ is the lattice constant, and
$c_{\mathbf{j}}=(c_{\mathbf{j}\uparrow}, c_{\mathbf{j}\downarrow})$ is
the annihilation operator at site $\mathbf{j}$ with spin $\uparrow,\downarrow$.
$\mathbf{j}=(\mathbf{j}_x,\mathbf{j}_z)$ in Eq.(\ref{eq:2}) but $\mathbf{j}=(\mathbf{j}_x,\mathbf{j}_y)$ in Eq.(\ref{eq:3}).
The effect of perpendicular magnetic field is included by
adding a phase term $\phi^{\perp}_{m,n} = -\int_{r_{m}}^{r_{n}}{\bf A(r)}\cdot d{\bf r}/\phi_0$,
with the vector potential ${\bf A}=(B_{y}z-B_{z}y,0,B_{x}y)$ and the flux quanta $\phi_0=\hbar/e$.
In the numerical calculations, we set the lattice constant $a =1.0\ \mathrm{nm}$,
and the Fermi velocity $v = 3.09\times10^{5}\ \rm m/s$.\cite{PengL}
The magnetic field is expressed in terms of the lattice magnetic flux $\phi$ with $\phi= B a^2/ 2\pi \phi_0$.
While $\phi=0.005$, the magnetic field $B$ is about $20.7$ Tesla.
The size of center region is $W\times L = 40a\times 8a$ [see the grey region in Fig.1(b)].
For the samples of other sizes, the conclusions are similar.

Considering a small temperature gradient $\Delta\mathcal{T}$ and
a zero bias applied on the longitudinal lead-1 and lead-2,
we can set the temperatures $\mathcal{T}_1=\mathcal{T}+\Delta\mathcal{T}/2$, $\mathcal{T}_2=\mathcal{T}-\Delta\mathcal{T}/2$, and $\mathcal{T}_3=\mathcal{T}_4=\mathcal{T}$,
and the biases $\mathrm{V}_q=0$ ($q=1, 2, 3, 4$), as shown in Fig.1(b).
Under the drive of the temperature gradient,
the spin current is induced in the transverse lead-3 and lead-4.
In usual, the spin current is a tensor and it has $3 \times 3=9$ elements
that respectively describe the direction of the electron motion
and the direction of the spin.\cite{spinc1,spinc2,spinc3}
While in the lead, the flow direction has to be along the lead direction,
but the spin direction may still be in the $x$, $y$ and $z$ directions.
So the spin current in the transverse lead may has three non-zero elements,
$I_{\mathrm{q}sx}$, $I_{\mathrm{q}sy}$ and $I_{\mathrm{q}sz}$. Here $I_{\mathrm{q}s\mathrm{i}}$ ($\mathrm{i}=x$, $y$ and $z$) represents an electron moving along the lead-$\mathrm{q}$ with its spin in the $\mathrm{i}$ direction.
Below we first derive the expression of the particle current $J_{\mathrm{qi\sigma}}$
($\sigma=\uparrow, \downarrow$ or $\sigma=+, -$),
which describes the particle current with the spin pointing the $\sigma \mathrm{i}$ direction in the lead-$\mathrm{q}$.
From the Landauer-B\"{u}ttiker formula, the particle current $J_{\mathrm{qi\sigma}}$
can be expressed as,\cite{ChengS,LiuX}
\begin{eqnarray}\label{eq:4}
J_{\mathrm{qi\sigma}}=\frac{1}{\hbar}\sum_{\mathrm{p}\neq \mathrm{q}}\sum_{k_y}\int T_{\mathrm{qi\sigma,p}}(E,k_y)[f_{\mathrm{q}}(E)-f_{\mathrm{p}}(E)]\text{d}E,
\end{eqnarray}
where $T_{\mathrm{qi\sigma,p}}(E,k_y)$ is
the transmission coefficient for the incident carrier from the lead-$\mathrm{p}$
with momentum $k_y$ to the $\sigma \mathrm{i}$ mode at the lead-$\mathrm{q}$.
The expression of the particle current $J_{\mathrm{qi\sigma}}$ in
Eq.(\ref{eq:4}) is obtained from the Hamiltonian in Eq.(\ref{eq:2}).
If for the Hamiltonian in Eq.(\ref{eq:3}), the sum over $k_y$ and $T_{\mathrm{qi\sigma,p}}(E,k_y)$
should be replaced by the sum over $k_z$ and $T_{\mathrm{qi\sigma,p}}(E,k_z)$.
Hereafter, we assume that the transverse lead-3 and lead-4 are normal conductance
without spin-orbit coupling. So the spin in the transverse leads is a good quantum number
and the $\sigma \mathrm{i}$ mode represents the transport mode with its spin pointing to $\sigma \mathrm{i}$ direction.
On the other hand, we set that the lead-1 and lead-2 are
the semi-infinite WSM leads which are the same as the rectangular center region.
That is to say, the lead-1/center WSM region/lead-2 forms a perfect WSM nanowire.
In Eq.(\ref{eq:4}),
\begin{eqnarray}\label{eq:6}
&f_{\mathrm{q}}(E,\mu_{\mathrm{q}},\mathcal{T}_{\mathrm{q}})=\frac{1}{e^{(E-\mu_{\mathrm{q}})/k_{B} \mathcal{T}_{\mathrm{q}}}+1}
\end{eqnarray}
is the electronic Fermi distribution function of the lead-$\mathrm{q}$, where the chemical potential $\mu_{\mathrm{q}} =E_F+eV_{\mathrm{q}}$ with the Fermi energy $E_F$.\cite{YangNX} $k_{B}$ is the Boltzmann constant.

By using the nonequilibrium Green's function method,
the transmission coefficient $T_{\mathrm{qi\sigma,p}}(E,k_y)$ can be obtained as:
\begin{eqnarray}\label{eq:add}
T_{\mathrm{qi\sigma,p}}(E,k_y)=\textmd{Tr}[{\bf \Gamma}_{\mathrm{qi\sigma}}
{\bf G}^r{\bf \Gamma}_{\mathrm{p}}{\bf G}^a],
\end{eqnarray}
in which
${\bf \Gamma}_{\mathrm{qi\sigma}}(E)=i[{\bf \Sigma}_{\mathrm{qi\sigma}}^r(E)
-{\bf \Sigma}_{\mathrm{qi\sigma}}^{r \dag}(E)]$ and
${\bf \Gamma}_{\mathrm{p}}(E)=i[{\bf \Sigma}_{\mathrm{p}}^r(E)
-{\bf \Sigma}_{\mathrm{p}}^{r \dag}(E)]$ are the linewidth functions.
${\bf \Sigma}_{\mathrm{qi\sigma}}^r$ and ${\bf \Sigma}_{\mathrm{p}}^r$
are the retarded self-energy due to the coupling between the lead and the center WSM region.
In Eq.(\ref{eq:add}), the Green's function
${\bf G}^r(E)=[{\bf G}^a]^{\dagger}=[E {\bf I}-
{\bf H}_{\mathrm{0}}-{\bf \Sigma}_{1}^r-{\bf \Sigma}_{2}^r-{\bf \Sigma}_{3}^r-{\bf \Sigma}_{4}^r]^{-1}$,
with ${\bf H}_{\mathrm{0}}$ being the Hamiltonian of center scattering region.\cite{Xingy,Longw}

For the normal transverse lead-3 and lead-4, the self-energy functions are
${\bf \Sigma}_{\mathrm{qi\sigma}}^r = -\frac{i\Gamma}{2} \bf{I}_{N} \otimes
{\bf u}_{\mathrm{i}}^{\dagger} \bf{m}_{\sigma} {\bf u}_{\mathrm{i}}$,
where $\Gamma$ describes the coupling strength between lead-$\mathrm{q}$ and the central region.
In the numerical calculation, we consider the coupling strength with $\Gamma =5.0$.
$\bf{m}_{\uparrow} =\left(
  \begin{array}{cc} 1 & 0 \\
 0 & 0 \\   \end{array} \right)$ for $\sigma=\uparrow$
and $\bf{m}_{\downarrow} =\left(
  \begin{array}{cc} 0 & 0 \\
 0 & 1 \\   \end{array} \right)$ for $\sigma=\downarrow$.
${\bf u}_{\mathrm{i}}$ is the unitary matrix which rotates the spin coordinate
with the spin $z$ axis being rotated into $\mathrm{i}$ direction.
Specifically, the unitary matrix ${\bf u}_{\mathrm{i}}$ ($\mathrm{i}=x,y,z$) is
\begin{eqnarray}\label{eq:5}
{\bf u}_{\mathrm{x}}&=&
\frac{\sqrt2}{2} \left(
  \begin{array}{cc}
    1 & 1 \\
    -1 & 1 \\
  \end{array}
\right),\\
{\bf u}_{\mathrm{y}}&=&
\frac{\sqrt2}{2} \left(
  \begin{array}{cc}
    i & 1 \\
    -i & 1 \\
  \end{array}
\right),\\
{\bf u}_{\mathrm{z}}&=&
\left(
  \begin{array}{cc}
    1 & 0\\
0 & 1 \\
  \end{array}
\right).
\end{eqnarray}
For the lead-3 and lead-4, the self-energy functions
${\bf \Sigma}_{\mathrm{q}}^r = {\bf \Sigma}_{\mathrm{qz\uparrow}}^r +{\bf \Sigma}_{\mathrm{qz\downarrow}}^r
={\bf \Sigma}_{\mathrm{qy\uparrow}}^r +{\bf \Sigma}_{\mathrm{qy\downarrow}}^r
={\bf \Sigma}_{\mathrm{qx\uparrow}}^r +{\bf \Sigma}_{\mathrm{qx\downarrow}}^r$.
For the semi-infinite WSM lead-1 and lead-2, the self-energy functions ${\bf \Sigma}_{\mathrm{p}}^r$
can be calculated numerically.\cite{Xingy,LeeD}

After obtaining the particle current $J_{\mathrm{qi\sigma}}$,
the charge current in lead-$\mathrm{q}$ can be obtained as $I_{\mathrm{q} e}=e(J_{\mathrm{q}x\uparrow}+J_{\mathrm{q}x\downarrow}) =e(J_{\mathrm{q}y\uparrow}+J_{\mathrm{q}y\downarrow})
=e(J_{\mathrm{q}z\uparrow}+J_{\mathrm{q}z\downarrow})$
and the spin current is $I_{\mathrm{q}s\mathrm{i}}=
(\hbar/2)(J_{\mathrm{qi}\uparrow}-J_{\mathrm{qi}\downarrow})$, straightforwardly.
Note here the charge current has an element $I_{\mathrm{q}}$ only, but
the spin current has three non-zero elements, $I_{\mathrm{q}sx}$, $I_{\mathrm{q}sy}$ and $I_{\mathrm{q}sz}$,
because that the charge current is vector but the spin current is a tensor.\cite{spinc1,spinc2,spinc3}
The spin Nernst coefficients in the lead-3 and lead-4
are defined as $N_{s\mathrm{qi}}\equiv I_{\mathrm{q}s\mathrm{i}}/\Delta\mathcal{T}$.
Here, $N_{s\mathrm{qi}}$ with $\mathrm{q}=3,4$ and $\mathrm{i}=x, y, z$ denotes the spin current induced in the transverse
lead-3 and lead-4 with the spin direction in the $\mathrm{i}$-direction by the longitudinal thermal gradient.
At the small thermal gradient limits, the spin Nernst coefficients $N_{s\mathrm{qi}}$ can be reduced to\cite{ChengS,LiuX}
\begin{eqnarray}
&&N_{s\mathrm{qi}}(E_F,\phi)=\sum_{k_y} \tilde{N}_{s\mathrm{qi}}(E_F,\phi,k_y) \nonumber\\
&& =
  \frac{1}{4\pi}\sum_{k_y}\int(\Delta T_{\mathrm{qi},2}-\Delta T_{\mathrm{qi},1})\frac{E-E_{F}}{k_{B}\mathcal{T}^2}f_0(1-f_0)\text{d}E,\nonumber\\
\label{eq:7}
\end{eqnarray}
where $\Delta T_{\mathrm{qi},\mathrm{p}}=T_{\mathrm{qi}\uparrow,\mathrm{p}}-T_{\mathrm{qi}\downarrow,\mathrm{p}}$,
and $f_0$ is the Fermi distribution function at the zero thermal gradient and zero bias.

\section{\label{sec3}
the thermospin transport in the $x$-$z$ mode}

First, we study the spin Nernst coefficients $N_{s3z}$, $N_{s3x}$ and $N_{s3y}$
in the case of $x$-$z$ transport mode.
In this $x$-$z$ mode, the thermal gradient is applied in the $x$ direction,
the spin current is measured at the $z$ direction, and
the magnetic field is in the $y$ direction which is perpendicular to the transport $x$-$z$ plane.
Fig.2(a) and Fig.3(a) show $N_{s3z}$ and $N_{s3x}$ as
functions of Fermi energy $E_F$ for different temperatures at the zero magnetic field.
When the magnetic field $B = 0$, the spin Nernst coefficient $N_{s3z}=0$ exactly
regardless of the temperature, but $N_{s3x}\neq 0$ and $N_{s3y}\neq 0$.
$N_{s3y}$ is not shown here, and it is non-zero but is much smaller than the $N_{s3x}$.
$N_{s3x}$ and $N_{s3y}$ exhibit a series of peaks at the low temperatures.
With the increase of temperature $\mathcal{T}$, $N_{s3x}$ and $N_{s3y}$ increase as a whole,
then some oscillation peaks merge.
The oscillation peaks of $N_{s3x}$ and $N_{s3y}$
become sparser at the higher temperature, and the peak spacings
of $N_{s3x}$ and $N_{s3y}$ become larger [see Fig.3(a)].
Besides, one can see that $N_{s3x}$ and $N_{s3y}$ are odd functions of
the Fermi energy $E_F$ with $N_{s3x/y}(-E_F)=-N_{s3x/y}(E_F)$.

\begin{figure}
\includegraphics[scale=0.33]{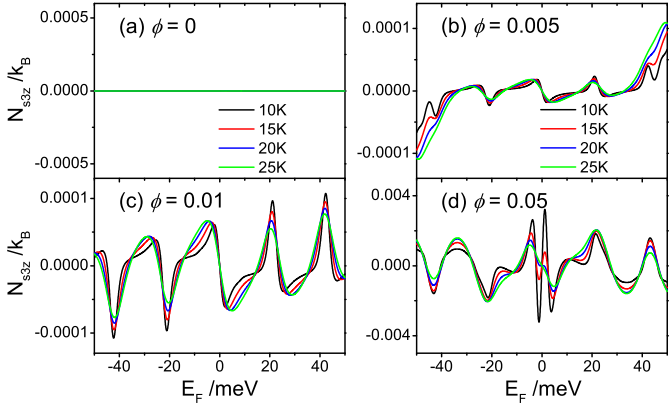}
\caption{ The $z$ direction element of spin Nernst coefficient, $N_{s3z}$,
versus Fermi energy $E_F$ for different temperatures in the $x$-$z$ mode,
in which the thermal gradient $\Delta\mathcal{T}$ is added in the $x$ direction
and the spin current is measured at the $z$ direction.
The size of the center WSM region is $40a \times 8a$ with $a =1.0\ \mathrm{nm}$
and the sum over $k_y$ is from $-\pi/a$ to $\pi/a$ with the interval $\pi/(30a)$,
which parameters are the same for all figures.
The perpendicular magnetic field $\phi =0$ (a), $\phi =0.005$ (b),
$\phi =0.01$ (c) and $\phi =0.05$ (d), respectively.  }
\end{figure}

\begin{figure}
\includegraphics[scale=0.33]{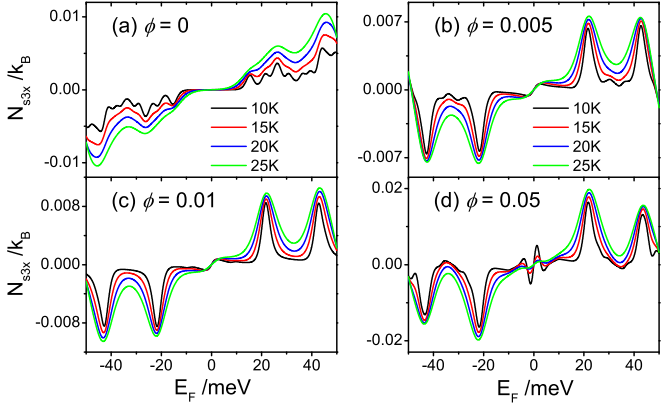}
\caption{ The $x$ direction element of spin Nernst coefficient, $N_{s3x}$,
versus Fermi energy $E_F$ for different temperatures in the $x$-$z$ mode.
The perpendicular magnetic field $\phi =0$ (a), $\phi =0.005$ (b), $\phi =0.01$ (c)
and $\phi =0.05$ (d), respectively.  }
\end{figure}

In order to explain the cause of the non-zero spin Nernst coefficient $N_{s3x}$,
we plot the transmission coefficient $T_{3x\sigma,p}$ at the momentum $k_y =0$
and the momentum resolved spin Nernst coefficient $\tilde{N}_{s3x}$ at several specific
momentum $k_{y}a =0$ and $\pm 0.1$ in Fig.4(a) and 4(b).
While the energy $E$ just crosses discrete transverse channels,
the transmission coefficients suddenly jump and show a peak.
At $k_{y} =0$ and $\phi =0$,
$T_{3x\uparrow,1}(E)=T_{3x\downarrow,2}(E)$ and $T_{3x\downarrow,1}(E)=T_{3x\uparrow,2}(E)$ exactly
[see Fig.4(a)], but $T_{3x\uparrow,1}(E)$ is not equal to $T_{3x\downarrow,1}(E)$.
This leads that $\Delta T_{3x,2} -\Delta T_{3x,1}$ has large non-zero value, so does
the spin Nernst coefficient $\tilde{N}_{s3x}$ [see Fig.4(b)].
However, $\Delta T_{3z,2} -\Delta T_{3z,1}$ is exactly zero at $k_{y}=0$ and $\phi = 0$,
which leads $\tilde{N}_{s3z} =0$.
When the momentum $k_y \not= 0$, Fig.4(b) exhibits that $\tilde{N}_{s3x}$ is an odd function
of Fermi energy $E_F$, but it is an even function of the momentum $k_y$ with
$\tilde{N}_{s3x}(k_{y})=\tilde{N}_{s3x}(-k_{y})$.

Let us explain why $N_{s3z}$ is zero, while $N_{s3x}$ has finite values at the
zero perpendicular magnetic field with aids of the physical picture in Fig.5.
In Eq.(1), the two Weyl cones of opposite chirality in WSMs are located at $\textbf{K}_{-}=(0,0,-\frac{\pi}{2a})$ and $\textbf{K}_{+}=(0,0,+\frac{\pi}{2a})$, as shown in Fig.5(a).
The Weyl Hamiltonian with $k_{y} = 0$ near two Weyl nodes $\textbf{K}_{\pm}$ can be approximated
as $\textbf{H}_{\pm}=\hbar v(k_{x}\sigma_{x}\mp k_{z}\sigma_{z})$.
The Fermi surface ($E_F \neq 0$) of these two Weyl cones is a circle as can be seen in Fig.5(b).
In particular, the spin direction and momentum is locked.
For a given momentum $\textbf{k}=(k_x, k_z)$,
the spin expectation value of the state $\Psi_{\textbf{k}\pm}$
is $\overrightarrow{s}_{\textbf{k} +} =\langle \Psi_{\textbf{k}+}|
(\sigma_x,\sigma_y,\sigma_z)|\Psi_{\textbf{k}+}\rangle
=(-\sin{\theta},0, -\cos{\theta})$ at one Weyl cone
($\theta$ is the angle between $\textbf{k}$ and $k_{z}$),
and it is $\overrightarrow{s}_{\textbf{k} -} = (-\sin{\theta}, 0, \cos{\theta})$ at other Weyl cone.
In Fig.5(b), the red arrow is the spin polarization direction of the state $\Psi_{\textbf{k}\pm}$.
Let us focus on the spin Nernst coefficients $N_{s3x}$ and $N_{s3z}$
and consider that the carriers are scattered from the lead-1 to the lead-3.
When the incident carriers are from the lead-1 with along $+x$ direction,
the state with positive momentum $k_x$ contributes the transmission coefficients
$T_{3x\sigma,1}$ and $T_{3z\sigma,1}$.
From Fig.5(b), one can see the $z$ direction elements of the spin polarization of states
$\Psi_{\textbf{k}\pm}$ at the two opposite-chirality Weyl cones are just opposite,
but the $x$ direction elements are the same.
So the $z$ direction element of spin Nernst coefficients ($N_{s3z}$)
contributed by the two opposite-chirality Weyl cones just cancel each other,
but the $x$ direction element ($N_{s3x}$) contributed by the two Weyl cones increases.
Furthermore, for all positive $k_x$ states, the $x$ direction elements of
the spin polarization point to the $-x$ direction, which leads a large negative value of $N_{s3x}$
at the negative Fermi energy $E_F$ as shown in Fig.3(a).

Next, we focus on the case of the non-zero perpendicular magnetic field.
Fig.2(b-d) and Fig.3(b-d) show the spin Nernst coefficients $N_{s3z}$ and $N_{s3x}$
versus the Fermi energy $E_F$
at the magnetic flux $\phi =0.005$, $\phi =0.01$ and $\phi =0.05$.
In the presence of the magnetic field, the $z$ direction element of
the spin Nernst coefficient, $N_{s3z}$, also appears.
In the traditional spin Nernst effect, the transverse spin current is induced
by the spin-orbit coupling.\cite{ChengS}
Here although there exists the spin-orbit coupling in the WSM,
the spin Nernst coefficient $N_{s3z}$ is zero at the zero magnetic field
and $N_{s3z}$ appears at the finite magnetic field.
This seems that the spin Nernst effect ($N_{s3z}$) is caused by the magnetic field,
not by the spin-orbit coupling.
So we call it the anomalous spin Nernst effect.
With the increase of $\phi$, $|N_{s3z}|$ increases.
But the $x$ direction element, $N_{s3x}$, is not increasing monotonously.
$N_{s3x}$ at the non-zero magnetic field can be smaller or larger than the value at the zero magnetic field.
In addition, $N_{s3x}$ exhibits a series of positive (negative) peaks while $E_F>0$ ($E_F<0$)
and $N_{s3z}$ oscillates between the positive and negative values again and again.
With the increase of temperature, the peak height of $N_{s3x}$ and oscillation amplitude of $N_{s3z}$
remain approximately unchanged, but the valley rises.
The $y$ direction element, $N_{s3y}$, has the similar behavior as $N_{s3x}$,
but its value is slightly smaller than $N_{s3x}$.
Furthermore, as shown in Figs.2 and 3, $N_{s3z}$, $N_{s3x}$ and $N_{s3y}$ are
all odd functions of the Fermi energy $E_F$.
The relational expression for positive and negative $E_F$ can be written as,
\begin{eqnarray}\label{eq:8}
N_{s3\mathrm{i}}(E_F,\phi)&=&-N_{s3\mathrm{i}}(-E_F,\phi), \quad \mathrm{i}=x,y,z.
\end{eqnarray}
On the other hand, the spin Nernst coefficient $N_{s3z}$ is an odd function of the magnetic field,
but $N_{s3x}$ and $N_{s3y}$ are even functions of the magnetic field, that is:
\begin{eqnarray}
N_{s3z}(E_F,\phi)&=&-N_{s3z}(E_F,-\phi), \label{eq:13}\\
N_{s3x/y}(E_F,\phi)&=&N_{s3x/y}(E_F,-\phi). \label{eq:13b}
\end{eqnarray}
From the relations of Eqs.(\ref{eq:13},\ref{eq:13b}), one can obtain that $N_{s3z}$ is zero
and $N_{s3x/y}$ can be finite at the zero magnetic field
straightforwardly.

\begin{figure}
\includegraphics[scale=0.33]{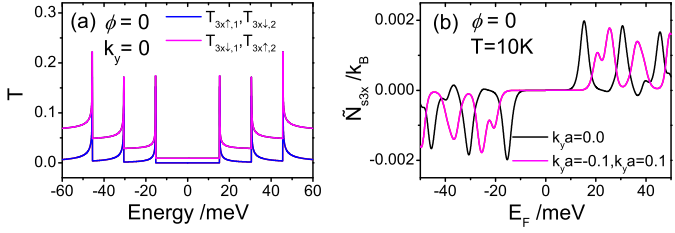}
\caption{
(a) shows the curves of the transmission coefficient $T_{3x\sigma,\mathrm{p}}$ versus
the incident energy $E$ with perpendicular magnetic field $\phi = 0$ and momentum $k_{y} =0$.
(b) is the momentum resolved spin Nernst coefficient $\tilde{N}_{s3x}$
versus the Fermi energy $E_F$ for $k_{y}a =0$, $-0.1$ and $0.1$
with $\phi =0$ and temperatures $\mathcal{T} =10K$.  }
\end{figure}

\begin{figure}
\includegraphics[scale=0.75]{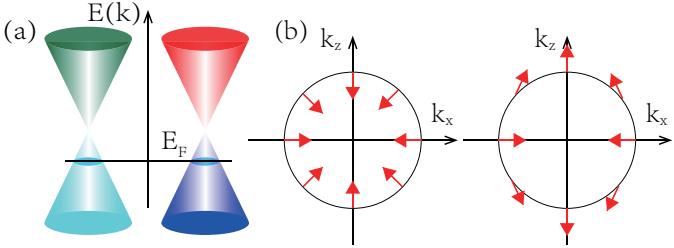}
\caption{
(a) Schematic illustrations for two opposite-chirality Weyl cones.
For a given Fermi energy, a round Fermi surface can be gotten from the Weyl cones.
(b) A sketch of the Fermi surface in the $k_x$-$k_z$ momentum space at $k_y =0$.
Here the two circles are the Fermi surfaces of the two Weyl cones with the opposite
chirality and these red arrows represent the spin polarization direction.}
\end{figure}

\begin{figure}
\includegraphics[scale=0.88]{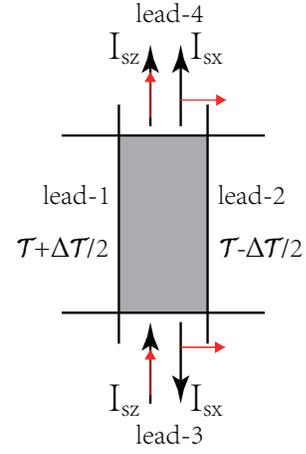}
\caption{
A sketch shows the spin polarization direction and the spin current direction
in the transverse lead-3 and lead-4 of the four-terminal crossbar sample.
Here the black arrows represent the spin current directions
and the red arrows are the spin polarization directions.
Here $I_{sz}$ flows in one lead and flows out on the other.
But $I_{sx}$ in the lead-3 and lead-4 flows out or flows in together. }
\end{figure}

Let us discuss the relationship of the spin Nernst coefficients at the transverse lead-3 and lead-4.
From numerical results, we obtain that $N_{s3\mathrm{i}}$ and $N_{s4\mathrm{i}}$ have the relations as:
\begin{eqnarray}
N_{s3z}(E_F,\phi)&=&-N_{s4z}(E_F,\phi),\label{eq:9}\\
N_{s3x/y}(E_F,\phi)&=& N_{s4x/y}(E_F,\phi).\label{eq:9b}
\end{eqnarray}
That is, under longitudinal thermal gradient drive,
the $z$ direction element of the spin current flows from the lead-3 through the
center WSM region to the lead-4, which is the same as the conventional spin Nernst effect.
However, the $x$ ($y$) direction element of the spin current
in the lead-3 and lead-4 flows out or flows in together,
in contrary to the traditional one.
So the $x$ and $y$ direction elements of the spin Nernst coefficient is also anomalous.
In Fig.6, we show the flowing direction and the spin polarization direction for the spin
currents $I_{sz}$ and $I_{sx}$ in the lead-3 and lead-4.
Here $I_{sx}$ in the lead-3 and lead-4 flows out or flows in together,
which are very different with the traditional spin Hall effect and spin Nernst effect.

These relations of the spin Nernst coefficients in Eqs.(\ref{eq:8}-\ref{eq:9b})
can also be obtained by analyzing the symmetry of the WSM Hamiltonian.
From the continuous Hamiltonian in Eq.(1) and the discrete Hamiltonian in Eq.(2),
we find that the WSM has three symmetries, the center-reversal-type symmetry, the
mirror-reversal-type symmetry and the electron-hole-type symmetry.

{\sl The center-reversal-type symmetry}:
In the continuous Hamiltonian in Eq.(1),
we take the transformation $(k_z, k_x, k_y)\rightarrow (-k_z, -k_x, -k_y)$ and rotate the spin by $180^{\circ}$
around spin $z$ axis (i.e. $\sigma_{z}\rightarrow \sigma_{z}$,
$\sigma_{x}\rightarrow -\sigma_{x}$ and $\sigma_{y}\rightarrow -\sigma_{y}$).
Or in the discrete Hamiltonian in Eq.(2), we take the transformation
$c_{{\bf j}\uparrow}\rightarrow ic_{-{\bf j}\uparrow}$,
$c_{{\bf j}\downarrow}\rightarrow -ic_{-{\bf j}\downarrow}$ and $k_y \rightarrow -k_y$.
Under this transformation the Hamiltonian $H$ is invariant,
but the lead-1 (lead-3) and lead-2 (lead-4) exchange each other.
So from this center-reversal-type symmetry, we can obtain
that the transmission coefficients have the following relations:
\begin{eqnarray}
T_{3z\sigma,1}(E_F,\phi,k_{y})&=&T_{4z\sigma,2}(E_F,\phi,-k_{y}), \label{eq:10a}\\
T_{3z\sigma,2}(E_F,\phi,k_{y})&=&T_{4z\sigma,1}(E_F,\phi,-k_{y}), \label{eq:10b} \\
T_{3x/y\sigma,1}(E_F,\phi,k_{y}) &=& T_{4x/y \bar{\sigma},2}(E_F,\phi,-k_{y}), \label{eq:10c} \\
T_{3x/y\sigma,2}(E_F,\phi,k_{y}) &=& T_{4x/y \bar{\sigma},1}(E_F,\phi,-k_{y}), \label{eq:10d}
\end{eqnarray}
where $\bar{\sigma}=\downarrow$ for $\sigma=\uparrow$ and $\bar{\sigma}=\uparrow$ for
$\sigma=\downarrow$.
Then, by combining Eqs.(\ref{eq:10a}-\ref{eq:10d}) and
Eq.(10), and summing over $k_{y}$, we can get
the relations in Eqs.(\ref{eq:9},\ref{eq:9b})
of the spin Nernst coefficients in lead-3 and lead-4, straightforwardly.

{\sl The mirror-reversal-type symmetry}:
If in the continuous Hamiltonian in Eq.(1),
we take $k_{z}\rightarrow -k_{z}$ and the magnetic flux $\phi\rightarrow -\phi$, or
if in the discrete Hamiltonian in Eq.(2), we take
$c_{j_x,j_z,\sigma}\rightarrow c_{j_x,-j_z,\sigma}$ and $\phi\rightarrow -\phi$,
the Hamiltonian $H$ remains the same.
Under this transformation, the lead-3 and lead-4 exchange each other.
So from this mirror-reversal-type symmetry, we have:
\begin{eqnarray}\label{eq:11}
T_{3x/y/z\sigma,\mathrm{p}}(E_F,\phi,k_y)=T_{4x/y/z\sigma,\mathrm{p}}(E_F,-\phi,k_y),
\end{eqnarray}
where $\mathrm{p}=1, 2$.
Then to substitute these relations of the transmission coefficients into
the expressions of $N_{s\mathrm{qi}}$ in Eq.(10),
one can get the relation between the spin Nernst coefficients $N_{s3\mathrm{i}}$ and $N_{s4\mathrm{i}}$
of the lead-3 and lead-4,
\begin{eqnarray}\label{eq:12ss}
N_{s3\mathrm{i}}(E_F,\phi)=N_{s4\mathrm{i}}(E_F,-\phi), \quad \mathrm{i}=x,y,z.
\end{eqnarray}
By combining with Eq.(\ref{eq:12ss}) and Eqs.(\ref{eq:9},\ref{eq:9b}),
we draw the relations in Eqs.(\ref{eq:13},\ref{eq:13b}),
that is, the spin Nernst coefficient $N_{s3z}$ is an odd function of the magnetic flux $\phi$
but $N_{s3x}$ and $N_{s3y}$ are even functions of $\phi$.

{\sl The electron-hole-type symmetry}:
In the discrete Hamiltonian in Eq.(2), if we take the transformation:
$c_{{\bf j}\uparrow}\rightarrow \tilde{c}_{{\bf j}\downarrow}^{\dag}$,
$c_{{\bf j}\downarrow}\rightarrow \tilde{c}_{{\bf j}\uparrow}^{\dag}$,
$k_{y}\rightarrow -k_{y}$ and $\phi \rightarrow -\phi$,
the Hamiltonian $H$ is invariant.
In this transformation, the electron annihilation operator is changed into the hole annihilation operator,
and so it is an electron-hole-type transformation and the energy $E$ will change into $-E$.
Due to the electron-hole-type symmetry, we can get the relations of the transmission coefficients:
\begin{eqnarray}
T_{3z\sigma,\mathrm{p}}(E,\phi,k_{y})&=&T_{3z\bar{\sigma},\mathrm{p}}(-E,-\phi,-k_{y}), \label{eq:14a} \\
T_{3x\sigma,\mathrm{p}}(E,\phi,k_{y})&=&T_{3x\sigma,\mathrm{p}}(-E,-\phi,-k_{y}),  \label{eq:14b}\\
T_{3y\sigma,\mathrm{p}}(E,\phi,k_{y})&=&T_{3y\sigma,\mathrm{p}}(-E,-\phi,-k_{y}).\label{eq:14c}
\end{eqnarray}
To substitute these relations of $T_{3x/y/z\sigma,\mathrm{p}}$ into the expressions of $N_{s\mathrm{qi}}$ in Eq.(10),
one can obtain the relations of the spin Nernst coefficients between positive and negative Fermi energy:
\begin{eqnarray}\label{eq:12}
N_{s3z}(E_F,\phi)&=&N_{s3z}(-E_F,-\phi), \\
N_{s3x/y}(E_F,\phi)&=&-N_{s3x/y}(-E_F,-\phi).
\end{eqnarray}
Then by combining with Eq.(\ref{eq:13},\ref{eq:13b}),
we have $N_{s3\mathrm{i}}(E_F,\phi)=-N_{s3\mathrm{i}}(-E_F,\phi)$ ($\mathrm{i}=x,y,z$) straightforwardly.
This means that the spin Nernst coefficients $N_{s3z}$, $N_{s3x}$ and $N_{s3y}$
are all odd functions of the Fermi energy $E_F$ regardless of other parameters
(e.g. temperature, magnetic flux, etc.),
which are completely consistent with the numerical results [see Figs.2 and 3].

\section{\label{sec4} the thermospin transport in the $z$-$x$ mode}

In this section, we study the thermospin transport behaviors in the $z$-$x$ transport mode.
In this $z$-$x$ connection mode, the longitudinal thermal gradient is applied in the $z$ direction,
the transverse spin current is measured at the $x$ direction, and
the magnetic field is still in the $y$ direction which is perpendicular to the transport $z$-$x$ plane.
The spin Nernst coefficients $N_{s3z}$ and $N_{s3y}$ versus the Fermi energy $E_F$
for the different magnetic field $\phi$ and different temperatures
are shown in Fig.7 and Fig.8, respectively.
When the magnetic field is absent with $\phi =0$, the spin Nernst coefficients
$N_{s3z}$, $N_{s3x}$ and $N_{s3y}$ are all zero.
In particular, $N_{s3z} =N_{s3x} =N_{s3y} =0$ can remain regardless of the temperature,
the size of the center WSM region, and Fermi energy.
This indicates that there is no spin Nernst effect in the $z$-$x$ transport mode
at the zero magnetic field, although there exists the spin-orbit coupling in the WSM.

No spin Nernst effect at the zero magnetic field can be explained
with aids of the momentum-spin locking bands as shown in Fig.5(b).
As mentioned in Sec.~\ref{sec3}, for a given momentum $\textbf{k}=(k_x, k_z)$,
the spin expectation value of the state $\Psi_{\textbf{k}\pm}$
is $\overrightarrow{s}_{\textbf{k} +} = (-\sin{\theta},0, -\cos{\theta})$ at one Weyl cone
and $\overrightarrow{s}_{\textbf{k} -} = (-\sin{\theta}, 0, \cos{\theta})$ at the other Weyl cone,
with $\theta$ being the angle between $\textbf{k}$ and $k_{z}$ [see Fig.5(b)].
In the $z$-$x$ mode, the lead-1 and lead-2 are along the $z$ direction.
Let us consider the incident carriers from the lead-1 which is along $+z$ direction.
So the state with positive momentum $k_z$ contributes the transmission coefficients
and the spin Nernst coefficients.
From Fig.5(b), one can see the $x$ direction elements of the spin polarization for
the $+k_x$ and $-k_x$ states ($\Psi_{(+k_x,k_z)\pm}$ and $\Psi_{(-k_x,k_z)\pm}$) are just opposite,
so the $x$ direction element of spin Nernst coefficients, $N_{s3x}$,
contributed by the $+k_x$ and $-k_x$ states cancel each other, leading
to $N_{s3x} =0$ regardless of the parameters of the system.
On the other hand, the $z$ direction elements of the spin polarization of states
$\Psi_{\textbf{k}\pm}$ at the two opposite-chirality Weyl cones exactly are opposite,
so the $z$ direction element of spin Nernst coefficients
contributed by the two opposite-chirality Weyl cones cancel each other also,
so that $N_{s3z} =0$.

When the perpendicular magnetic field $\phi$ along the $y$ direction is applied,
the spin Nernst coefficients $N_{s3z}$, $N_{s3x}$ and $N_{s3y}$ appear with the finite values
[see Fig.7(b-d) and Fig.8(b-d)].
With the increase of $\phi$, the spin Nernst coefficients increase rapidly at the beginning,
then slowly, but generally keeps increasing.
These results seem to indicate that the spin Nernst effect is
caused by magnetic field, not by the spin-orbit coupling.
So it is an anomalous spin Nernst effect and is essentially different
with the traditional spin Nernst effect which can appear
at the zero magnetic field.
This anomalous spin Nernst effect originates from the
combination of the magnetic field and the peculiar band structure of the WSMs.
Here the spin Nernst coefficients $N_{s3z}$ and $N_{s3y}$ show some peaks [see Figs.7 and 8].
$N_{s3z}$ has the positive peaks while $E_F>0$ and the negative peaks at $E_F<0$.
All peaks of $N_{s3y}$ are negative regardless of the value of Fermi energy $E_F$.
While $E_F$ is near the Weyl nodes, $N_{s3y}$ has the highest peak
but $N_{s3z}$ is very small. When the temperature $\mathcal{T}$ rises,
the height and position of the peaks of $N_{s3z}$ and $N_{s3y}$ roughly remain unchanged,
and the valleys of $N_{s3z}$ rise but the valleys of $N_{s3y}$ do not change almost.
When the magnetic field $\phi$ increases further from the moderate value
(e.g. $\phi =0.005$), the peak heights of $N_{s3z}$ and $N_{s3y}$ slightly increase,
but the peak positions remain almost unchanged.
In particular, the details of the curves of $N_{s3y}$-$E_F$ for
the different $\phi$ are very similar [see Fig.8(b-d)].
Here $N_{s3x}$ is not shown. $N_{s3x}$ and $N_{s3y}$ have similar characteristics
and they are about the same value.

\begin{figure}
\includegraphics[scale=0.33]{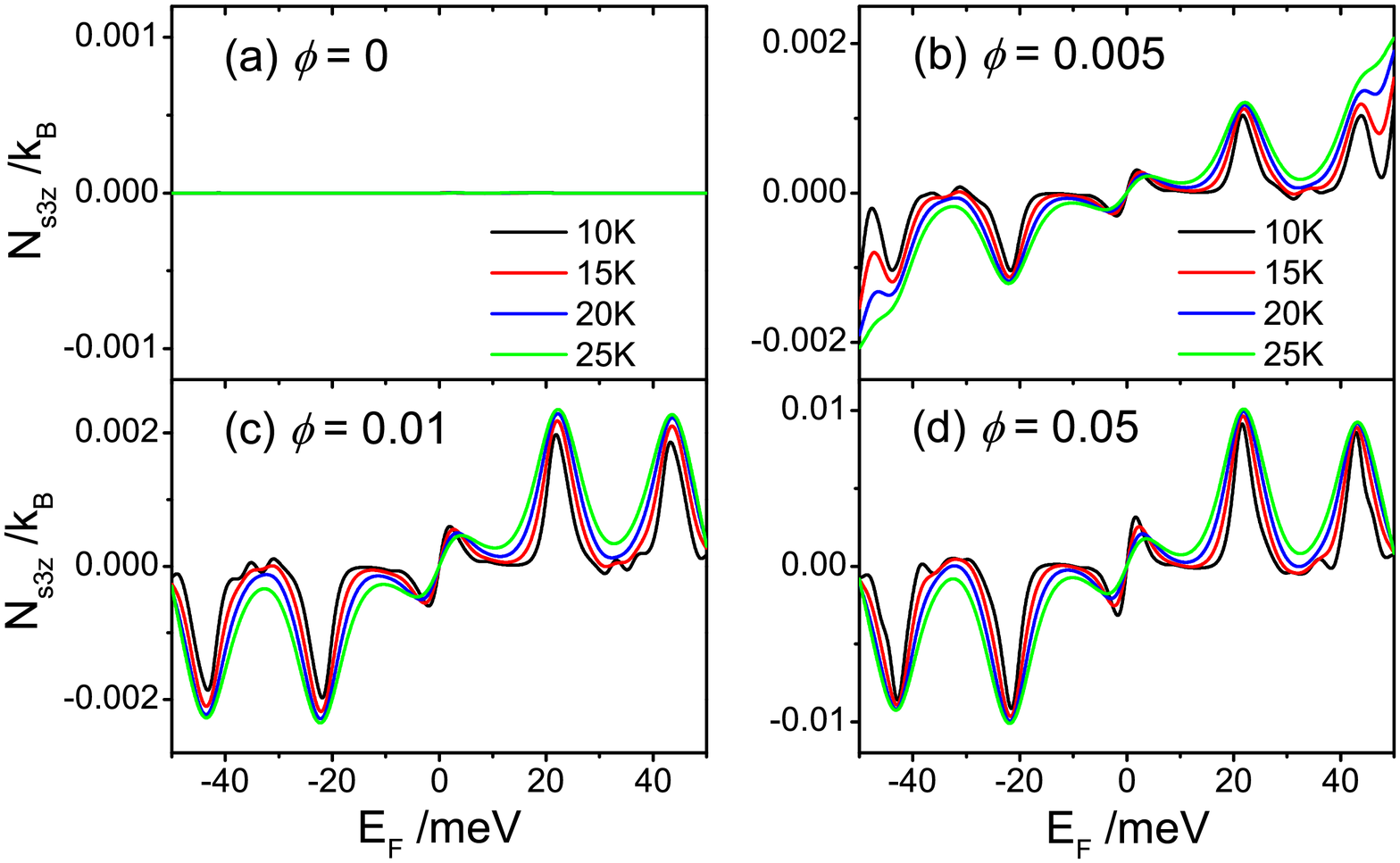}
\caption{
The $z$ direction element of spin Nernst coefficient, $N_{s3z}$,
versus Fermi energy $E_F$ for different temperatures in the $x$-$z$ mode,
in which the thermal gradient $\Delta\mathcal{T}$ is added in the $z$ direction
and the spin current is measured at the $x$ direction.
The perpendicular magnetic field $\phi =0$ (a), $\phi =0.005$ (b),
$\phi =0.01$ (c) and $\phi =0.05$ (d), respectively.  }
\end{figure}

\begin{figure}
\includegraphics[scale=0.33]{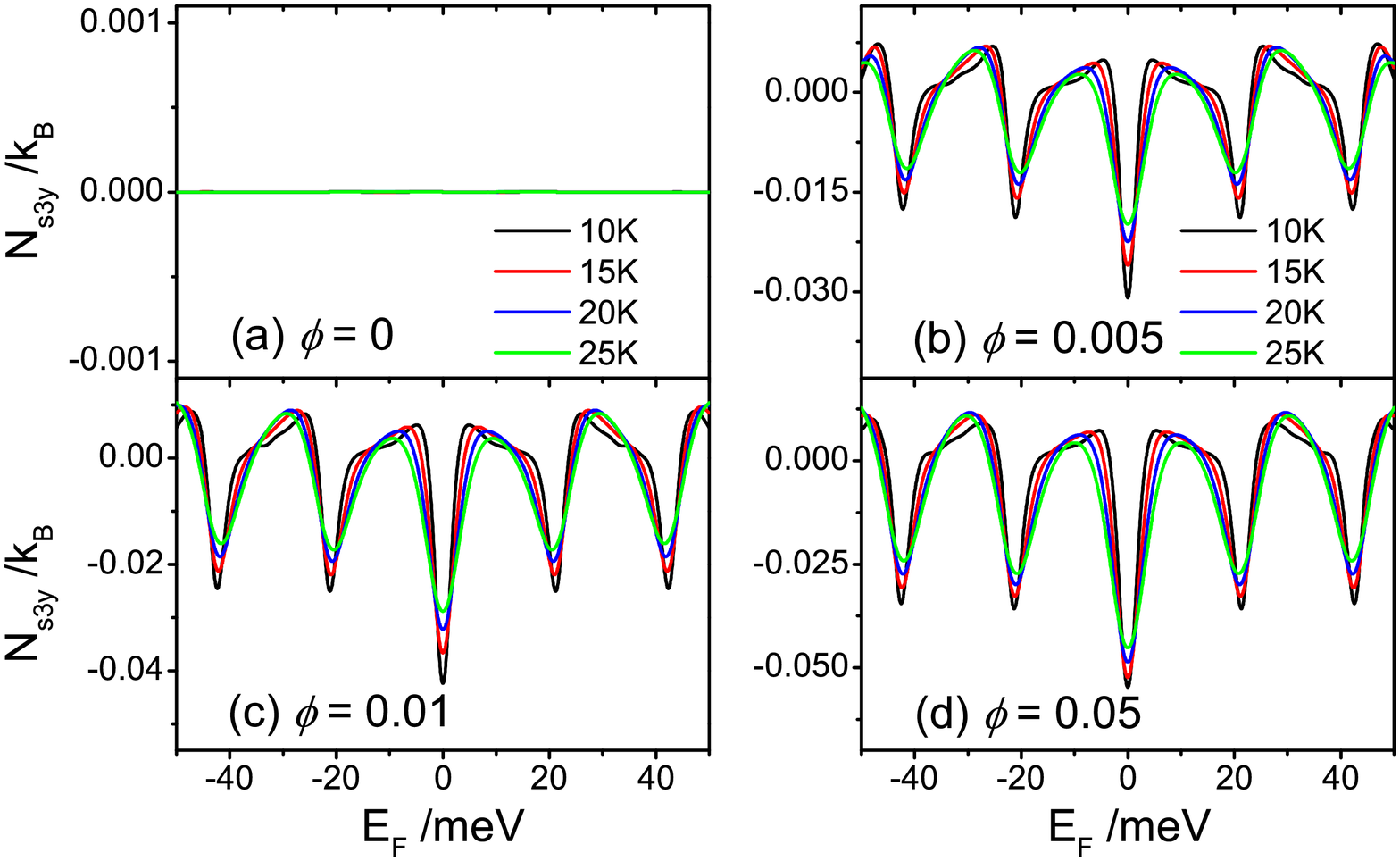}
\caption{
The $y$ direction element of spin Nernst coefficient, $N_{s3y}$,
versus Fermi energy $E_F$ for different temperatures in the $z$-$x$ mode.
The perpendicular magnetic field $\phi =0$ (a), $\phi =0.005$ (b), $\phi =0.01$ (c)
and $\phi =0.05$ (d), respectively. }
\end{figure}

In order to explain the origination of the peaks in the curves of the
spin Nernst coefficients versus Fermi energy, we show
the momentum resolved spin Nernst coefficient $\tilde{N}_{s3y}$
for the different magnetic fields in Fig.9.
Here each curve of $\tilde{N}_{s3y}$ shows a high peak. The peak position is
about at $E_F = 2ta k_y = \hbar v k_y$. For example, while $k_y=\pi/30 a$ and $v=3.09\times 10^5$m/s,
$\hbar v k_y \approx 21.3$meV, which is just the peak position of $\tilde{N}_{s3y}$ at $k_y =\pi/30 a$.
In fact, for a given momentum $k_y$, the WSMs in Hamiltonians (1) and (2)
reduce into the two-dimensional system,
and the Landau levels form under the strong magnetic field $B$.
The Landau levels are at $\pm\sqrt{(\hbar v k_y)^2 +2\hbar v^2 eB n}$ with the
level index $n=0,\pm1,\pm2,...$.
That is, that the peak position of the momentum resolved $\tilde{N}_{s3y}$ is just at
the zeroth Landau level.
Because that the zeroth Landau level is not almost affected by the magnetic field $B$,
the peak position as well the details of the curves $N_{s3\mathrm{i}}$-$E_F$ and
$\tilde{N}_{s3\mathrm{i}}$-$E_F$ in Fig.7, Fig.8 and Fig.9 are almost independent of $B$ also.
In addition, in the numerical calculations, we consider that
the thickness of the WSMs in $y$ direction is $61a$.
If we consider the thicker WSMs, the values of momentum $k_y$ are denser,
then the peaks at the curves of $N_{s3y}$-$E_F$ in Fig.8 are also denser.
For a very thick WSM, these peaks merge and cause large
spin Nernst coefficient $N_{s3y}$ over a wide $E_F$ range.

\begin{figure}
\includegraphics[scale=0.33]{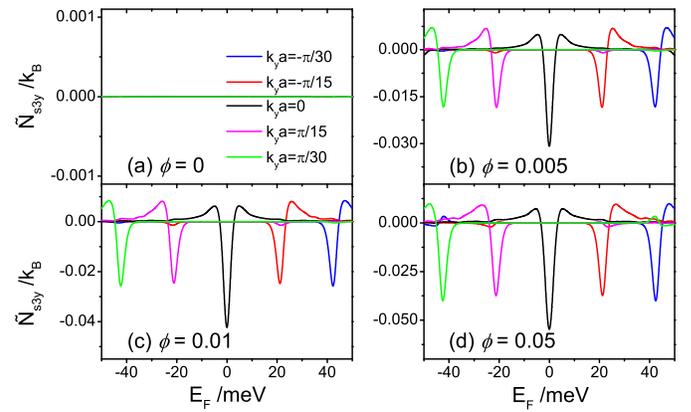}
\caption{
The momentum resolved spin Nernst coefficient $\tilde{N}_{s3y}$
versus the Fermi energy $E_F$ for $k_{y}a =0$, $\pm\pi/30$ and $\pm\pi/15$
with temperatures $\mathcal{T} =10K$ and the magnetic field $\phi =0$ (a),
$\phi =0.005$ (b), $\phi =0.01$ (c) and $\phi =0.05$ (d).  }
\end{figure}

In addition, from Figs.7 and 8, we can see that
the $z$ direction element of spin Nernst coefficient, $N_{s3z}$,
is an odd function of the Fermi energy $E_F$, but the $x$ and $y$ direction elements,
$N_{s3x}$ and $N_{s3y}$, are even functions of $E_F$. That is,
the spin Nernst coefficients have the following relations:
\begin{eqnarray}
N_{s3z}(E_F,\phi)=-N_{s3z}(-E_F,\phi), \label{eq:21a} \\
N_{s3x/y}(E_F,\phi)=N_{s3x/y}(-E_F,\phi).\label{eq:21b}
\end{eqnarray}
These relations are different with the $x$-$z$ mode, in which all elements of
the spin Nernst coefficients are odd functions of $E_F$ [see Eq.(\ref{eq:8})].

Moreover, the calculation results show all elements of
the spin Nernst coefficients are the odd functions of the
magnetic field $\phi$, i.e.
\begin{eqnarray}\label{eq:18}
N_{s3\mathrm{i}}(E_F, \phi)=-N_{s3\mathrm{i}}(E_F, -\phi), \quad \mathrm{i}=x,y,z.
\end{eqnarray}
From the relations of the odd functions of $\phi$,
we can obtain $N_{s3\mathrm{i}}(E_F) =0$ at the $\phi=0$ straightforwardly.

Let us study the relation of the spin Nernst coefficients at the transverse lead-3 and lead-4.
From calculation results, we obtain that $N_{s3\mathrm{i}}$ and $N_{s4\mathrm{i}}$ have the relations as:
\begin{eqnarray}
N_{s3z}(E_F,\phi)&=&-N_{s4z}(E_F,\phi),\label{eq:99}\\
N_{s3x/y}(E_F,\phi)&=& N_{s4x/y}(E_F,\phi).\label{eq:99b}
\end{eqnarray}
These relations are the same as that in the $x$-$z$ mode [see Eqs.(\ref{eq:9},\ref{eq:9b})].
That is, under longitudinal thermal gradient drive,
the $z$ direction element of the spin current, $I_{sz}$, flows
from a transverse lead through the center WSM region to the other transverse lead,
but the $x$ and $y$ direction elements, $I_{sx}$ and $I_{sy}$,
in two transverse leads flow out or flow in together, as shown in Fig.6.

In fact, these relations of the spin Nernst coefficients in Eqs.(\ref{eq:21a}-\ref{eq:99b})
obtained from the numerical results
can analytically be derived from the three symmetries mentioned above also.
First, from the center-reversal-type symmetry,
one can get that the transmission coefficients in the $z$-$x$ transport mode
have the relations as shown in Eqs.(\ref{eq:10a}-\ref{eq:10d}) still.
So the relations of spin Nernst coefficients $N_{s3\mathrm{i}}$ and $N_{s4\mathrm{i}}$
in the transverse lead-3 and lead-4 in Eqs.(\ref{eq:99},\ref{eq:99b})
can be obtained straightforwardly.

Second, in the mirror-reversal-type symmetry, the momentum $k_z \rightarrow -k_z$ and the
magnetic flux $\phi\rightarrow -\phi$. That is, the longitudinal lead-1 and lead-2 exchange each other.
Therefore from the mirror-reversal-type symmetry, we get the relations of the
transmission coefficients:
\begin{eqnarray}\label{eq:19}
T_{3x/y/z\sigma,1}(E,\phi, k_{y}) & = &
T_{3x/y/z\sigma,2}(E, -\phi, k_{y}).
\end{eqnarray}
To substitute these relations of $T_{3\mathrm{i}\sigma,\mathrm{p}}$ into the expression of the spin Nernst
coefficients in Eq.(\ref{eq:7}), one has:
\begin{eqnarray}\label{eq:201}
\tilde{N}_{s3x/y/z}(E_F,\phi, k_{y}) & = &
- \tilde{N}_{s3x/y/z}(E_F, -\phi, k_{y}).
\end{eqnarray}
Then by summing over the momentum $k_y$, the relations in Eq.(\ref{eq:18}),
the spin Nernst coefficients being the odd functions of the
magnetic field $\phi$, can be obtained straightforwardly.

Third, from the electron-hole-type symmetry,
we can get the relations of the transmission coefficients:
\begin{eqnarray}
T_{3z\sigma,\mathrm{p}}(E,\phi,k_{y}) &=&T_{3z\bar{\sigma},\mathrm{p}}(-E,-\phi,-k_{y}), \label{eq:202a} \\
T_{3x/y\sigma,\mathrm{p}}(E,\phi,k_{y})&=&T_{3x/y\sigma,\mathrm{p}}(-E,-\phi,-k_{y}),  \label{eq:202a}
\end{eqnarray}
which are the same as the relations in Eqs.(\ref{eq:14a}-\ref{eq:14c}) in the $x$-$z$ mode case.
Then to substitute these relations of $T_{3\mathrm{i}\sigma,\mathrm{p}}$ into Eq.(\ref{eq:7}), we obtain
\begin{eqnarray}
\tilde{N}_{s3z}(E_F,\phi,k_y)&=&\tilde{N}_{s3z}(-E_F,-\phi,-k_y), \label{eq:203a}\\
\tilde{N}_{s3x/y}(E_F,\phi,k_y)&=&-\tilde{N}_{s3x/y}(-E_F,-\phi,-k_y). \label{eq:203b}
\end{eqnarray}
To combine the above equations with the Eq.(\ref{eq:201}), we have:
\begin{eqnarray}
\tilde{N}_{s3z}(E_F,\phi,k_y)&=&-\tilde{N}_{s3z}(-E_F,\phi,-k_y), \label{eq:204a}\\
\tilde{N}_{s3x/y}(E_F,\phi,k_y)&=&\tilde{N}_{s3x/y}(-E_F,\phi,-k_y). \label{eq:204b}
\end{eqnarray}
The analytical results in Eq.(\ref{eq:204b}) obtained from systemic symmetry
completely agree with the numerical calculations in Fig.9.
At last, by summing over the momentum $k_y$ in Eqs.(\ref{eq:204a} and \ref{eq:204b}),
one get the relations in Eqs.(\ref{eq:21a} and \ref{eq:21b}), i.e.,
the $z$ direction element of spin Nernst coefficient being an odd function of the Fermi energy
but the $x$ and $y$ direction elements being even functions.
In short, all the relations [in Eqs.(\ref{eq:21a}-\ref{eq:99b})]
of the spin Nernst coefficients over the Fermi energy,
the magnetic field and the transverse terminals obtained from the numerical calculations
can also be derived by the systemic symmetry analysis.

\begin{figure}
\includegraphics[scale=0.33]{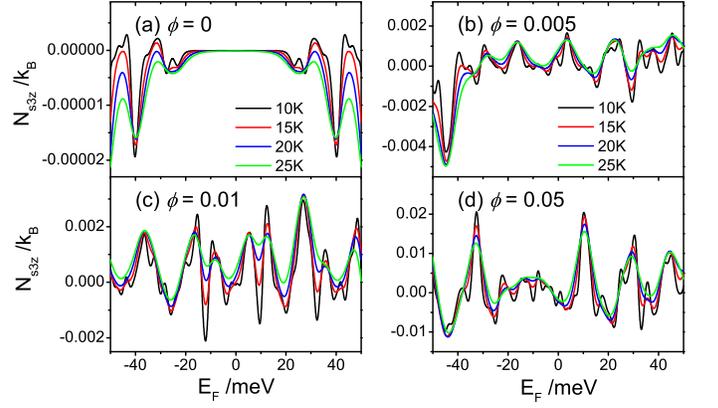}
\caption{
The $z$ direction element of spin Nernst coefficient, $N_{s3z}$,
versus $E_F$ for different temperatures in the $x$-$y$ mode.
Here the thermal gradient $\Delta\mathcal{T}$ is added in the $x$ direction,
the spin current is measured at the $y$ direction and the magnetic field
is applied at the $z$ direction.
The magnetic field $\phi =0$ (a), $\phi =0.005$ (b),
$\phi =0.01$ (c) and $\phi =0.05$ (d), respectively.  }
\end{figure}

\begin{figure}
\includegraphics[scale=0.33]{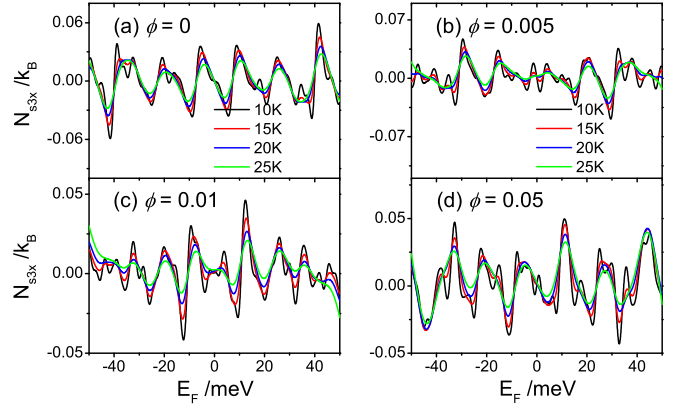}
\caption{
The $x$ direction element of spin Nernst coefficient, $N_{s3x}$,
versus Fermi energy $E_F$ for different temperatures in the $x$-$y$ mode.
The magnetic field $\phi =0$ (a), $\phi =0.005$ (b), $\phi =0.01$ (c)
and $\phi =0.05$ (d), respectively. }
\end{figure}

\section{\label{sec5} the thermospin transport in the $x$-$y$ mode}

In this section, we study the thermospin transport behaviors in the $x$-$y$ connection mode.
In this $x$-$y$ mode, the longitudinal thermal gradient is added in the $x$ direction,
the transverse spin current is measured at the $y$ direction, and
the magnetic field is in the $z$ direction as shown in Fig.2(b) and 2(e).
Figs.10(a) and 11(a) show the spin Nernst coefficients
$N_{s3z}$ and $N_{s3x}$ versus the Fermi Energy $E_F$ for different temperatures
at the zero magnetic field ($\phi =0$).
Here three elements $N_{s3z}$, $N_{s3x}$ and $N_{s3y}$ are all non-zero even if at $\phi=0$,
which are essentially different from that of the $x$-$z$ and $z$-$x$ modes.
$N_{s3z}$ is an even function of the Fermi energy with $N_{s3z}(E_F)=N_{s3z}(-E_F)$,
while $N_{s3x}$ and $N_{s3y}$ are odd functions with $N_{s3x/y}(E_F)=-N_{s3x/y}(-E_F)$.
$N_{s3z}$ shows some negative peaks but the value of $N_{s3z}$ is very small.
On the other hand, $N_{s3x}$ strongly oscillates between the positive and negative values.
The oscillation amplitude of $N_{s3x}$ is quite large and it can be over $0.05k_{B}$.
With the increase of temperature $\mathcal{T}$, the peak height of $N_{s3z}$ almost remains
and the oscillation amplitude of $N_{s3x}$ slightly reduces.
$N_{s3y}$ has the similar characteristics as $N_{s3x}$, but the value of $N_{s3y}$ is much smaller
than $N_{s3x}$.

Let us explain why $N_{s3x}$ has a large value and all three elements of spin Nernst coefficients
are non-zero in the $x$-$y$ connection mode.
In the case of the $x$-$y$ mode, the spin Nernst coefficients are mainly contributed
by these carriers with $k_z$ near $0$.
The Weyl Hamiltonian with $k_{z} =0$ can be approximated as $\textbf{H}_{\pm}=\hbar v(k_{x}\sigma_{x}+k_{y}\sigma_{y})$.
Here the Hamiltonian $\textbf{H}_{\pm}$ of two Weyl cones near $\textbf{K}_{\pm}$ are the same.
So the spin Nernst coefficients from two Weyl cones can not cancel always,
which is essentially different with the $x$-$z$ and $z$-$x$ modes as shown in Fig.5.
In addition, if we consider the incident carriers from the lead-1,
the states with the positive momentum $k_x$ contribute the spin Nernst coefficients.
Note for all positive $k_x$ states, the $x$ direction elements of
their spin polarization have the same sign, leading to a large value of $N_{s3x}$.

Next, we study the effect of the magnetic field on the spin Nernst effect.
Figs.10(b-d) and 11(b-d) show the spin Nernst coefficients $N_{s3z}$ and $N_{s3x}$
for the magnetic field $\phi =0.005$, $\phi =0.01$ and $\phi =0.05$, respectively.
With the increase of the magnetic field $\phi$, the $z$ direction element of
the spin Nernst coefficient, $N_{s3z}$, increases as a whole.
But it does not increase monotonously, e.g. $N_{s3z}$ at $\phi =0.01$ is smaller than
the value at $\phi =0.005$ [see Fig.10(b) and 10(c)].
The $x$ direction element, $N_{s3x}$, still keeps the large value and the strong
irregular oscillation in the presence of the magnetic field.
In addition, when $\phi \neq 0$, the spin Nernst coefficient $N_{s3z}$ is
both non-odd and non-even function of the Fermi energy $E_F$ and magnetic field $\phi$.
However, when the Fermi energy $E_F$ and the magnetic field $\phi$
change the sign at the same time, the spin Nernst coefficients have the relations:
\begin{eqnarray}\label{eq:27}
N_{s3z}(E_F,\phi) &=& N_{s3z}(-E_F,-\phi), \label{eq:27a}   \\
N_{s3x/y}(E_F,\phi)& =& -N_{s3x/y}(-E_F,-\phi).\label{eq:27b}
\end{eqnarray}
Furthermore, from the numerical results, we also get the relations of the spin Nernst coefficients
at the transverse lead-3 and lead-4:
\begin{eqnarray}
N_{s3z}(E_F,\phi)&=&-N_{s4z}(E_F,\phi), \label{eq:25a}\\
N_{s3x/y}(E_F,\phi)&=& N_{s4x/y}(E_F,\phi).\label{eq:25b}
\end{eqnarray}
That is, the $z$ direction element of the spin current, $I_{sz}$,
flows from a transverse lead through the center WSM region to the other transverse lead
which is the same with the conventional spin Nernst effect.
But the $x$ and $y$ direction elements, $I_{sx}$ and $I_{sy}$,
in two transverse leads flow out or flow in together, as shown in Fig.6,
which is abnormal.
These relations in Eqs.(\ref{eq:25a},\ref{eq:25b}) are the same as
that in the $x$-$z$ and $z$-$x$ modes.

Let us analytically derive the relations in Eqs.(\ref{eq:27a}-\ref{eq:25b})
from the systemic symmetry.
First, the mirror-reversal-type symmetry in the $x$-$y$ mode is slightly different
with that in the $x$-$z$ and $z$-$x$ modes, because of the difference of the direction of
the magnetic field.
In the present $x$-$y$ mode, we take the transformation of the momentum $k_z \rightarrow -k_z$
and the magnetic flux $\phi\rightarrow \phi$ (i.e. the magnetic field does not need an inverse sign),
the Hamiltonian $H$ is invariant.
From this mirror-reversal symmetry, although we can obtain
$ T_{3\mathrm{i}\sigma,1}(E_F,\phi,k_z)=T_{3\mathrm{i}\sigma,1}(E_F,\phi,-k_z)$
and $ T_{3\mathrm{i}\sigma,2}(E_F,\phi,k_z)=T_{3\mathrm{i}\sigma,2}(E_F,\phi,-k_z)$ with $\mathrm{i}=x,y,z$,
from these relations of the transmission coefficients one only gets $N_{s3\mathrm{i}}(E_F,\phi)=N_{s3\mathrm{i}}(E_F,\phi)$.

Second, from the center-reversal-type symmetry, we have
\begin{eqnarray}
T_{3z\sigma,1}(E_F,\phi,k_{z})&=&T_{4z\sigma,2}(E_F,\phi,-k_{z}), \label{eq:210a}\\
T_{3z\sigma,2}(E_F,\phi,k_{z})&=&T_{4z\sigma,1}(E_F,\phi,-k_{z}), \label{eq:210b} \\
T_{3x/y\sigma,1}(E_F,\phi,k_{z}) &=& T_{4x/y \bar{\sigma},2}(E_F,\phi,-k_{z}), \label{eq:210c} \\
T_{3x/y\sigma,2}(E_F,\phi,k_{z}) &=& T_{4x/y \bar{\sigma},1}(E_F,\phi,-k_{z}). \label{eq:210d}
\end{eqnarray}
These relations of the transmission coefficients are similar with
that in Eqs.(\ref{eq:10a}-\ref{eq:10d}) for the $z$-$x$ and $x$-$z$ modes.
To combine these relations and Eq.(\ref{eq:7}),
the relations of spin Nernst coefficients $N_{s3\mathrm{i}}$ and $N_{s4\mathrm{i}}$
in the two transverse lead-3 and lead-4 in Eqs.(\ref{eq:25a},\ref{eq:25b})
can be derived analytically.
From the above results, we find that the central-reversal-type symmetry
makes that the spin current $I_{sz}$
flows from a transverse lead to the other transverse lead
and the spin currents $I_{sx}$ and $I_{sy}$
in two transverse leads flow out or flow in together in all three
($x$-$z$, $z$-$x$ and $x$-$y$) connection modes.

Finally, from the electron-hole-type symmetry,
we can get the relations of the transmission coefficients:
\begin{eqnarray}
T_{3z\sigma,\mathrm{p}}(E,\phi,k_{z})=T_{3z\bar{\sigma},\mathrm{p}}(-E,-\phi,-k_{z}), \label{eq:214a} \\
T_{3x\sigma,\mathrm{p}}(E,\phi,k_{z})=T_{3x\sigma,\mathrm{p}}(-E,-\phi,-k_{z}),  \label{eq:214b}\\
T_{3y\sigma,\mathrm{p}}(E,\phi,k_{z})=T_{3y\sigma,\mathrm{p}}(-E,-\phi,-k_{z}).\label{eq:214c}
\end{eqnarray}
To substitute these relations of transmission coefficients into
the expressions of spin Nernst coefficients in Eq.(\ref{eq:7}),
one can analytically obtain the relations of the spin Nernst coefficients
in Eqs.(\ref{eq:27a},\ref{eq:27b}), straightforwardly.

\section{\label{sec6} Conclusions}

In summary, we study the spin Nernst effect for the Weyl semimetals
under the perpendicular magnetic fields.
By using nonequilibrium Green function method combining with the tight-binding Hamiltonian,
three elements of the spin currents and spin Nernst coefficients at the transverse leads are derived.
Due to the anisotropy of the Weyl semimetals,
the spin Nernst coefficients are strongly dependent on both the direction of thermal gradient and
the direction of the transverse lead connection.
There are three non-equivalent modes ($x$-$z$, $z$-$x$ and $x$-$y$ modes), which are studied in detail.
For the $\mathrm{i}$-$\mathrm{j}$ mode ($\mathrm{i},\mathrm{j}=x, y, z$), the thermal gradient is applied in the $\mathrm{i}$ direction and
the spin current is measured in the $\mathrm{j}$ direction.
We find that the spin Nernst effect in the Weyl semimetals is essentially different with
the traditional spin Nernst effect.
So we call it the anomalous spin Nernst effect.
Its anomalous behavior is manifested in the following two aspects.
1). The spin Nernst coefficients are zero at the zero magnetic field,
and they appear at the presence of the magnetic field.
This seems that the spin Nernst effect is caused by the magnetic field,
in contrary to the traditional one induced by the spin-orbit coupling.
2). The $x$ and $y$ direction elements of the spin currents in the two transverse leads
flows out or flows in together.
This is very different with the traditional one, where
the spin current flows out from one transverse lead and flows in on the other
transverse lead.

In addition, we show that the Weyl semimetals have the center-reversal-type symmetry, the
mirror-reversal-type symmetry and the electron-hole-type symmetry.
From the three symmetry, the spin Nernst coefficients are odd functions or even functions
of the Fermi energy, the magnetic field and the transverse terminals.
In particular, these odd or even relations of the spin Nernst coefficients
analytically obtained from symmetries
are completely consistent with the numerical calculation results.
Furthermore, the spin Nernst coefficients show the strong anisotropic characteristics.
At the zero magnetic field, the $z$ direction elements of the spin Nernst coefficients
are zero for the $x$-$z$ and $z$-$x$ modes, but it has non-zero value for the $x$-$y$ mode.
In the presence of the magnetic field,
for the $x$-$z$ and $z$-$x$ modes the spin Nernst coefficients show a series of peaks
and the peak positions are independent the magnetic field and temperatures.
But for the $x$-$y$ mode, the spin Nernst coefficients strongly oscillate
between the positive and negative values and are very sensitive to the magnetic field.
These strongly anisotropic behaviors of the spin Nernst effect
can be used as the characterization of magnetic Weyl semimetals
with broken time reversal symmetry.
It is also hoped that these anomalous behaviors of the spin Nernst effect
will be helpful to control the spin current.

\section*{Acknowledgement}
This work was financially supported by National Key R and D Program of China (Grant No. 2017YFA0303301),
NBRP of China (Grant No. 2015CB921102), NSF-China (Grant No. 11574007),
the Strategic Priority Research Program of Chinese Academy of Sciences (Grant No. XDB28000000),
and Beijing Municipal Science \& Technology Commission (Grant No.Z181100004218001).

\end{document}